\PassOptionsToPackage{unicode}{hyperref}
\PassOptionsToPackage{hyphens}{url}
\PassOptionsToPackage{dvipsnames,svgnames,x11names}{xcolor}
\documentclass[
]{article}
\usepackage{amsmath,amssymb}
\usepackage{lmodern}
\usepackage{iftex}
\ifPDFTeX
  \usepackage[T1]{fontenc}
  \usepackage[utf8]{inputenc}
  \usepackage{textcomp} 
\else 
  \usepackage{unicode-math}
  \defaultfontfeatures{Scale=MatchLowercase}
  \defaultfontfeatures[\rmfamily]{Ligatures=TeX,Scale=1}
\fi
\IfFileExists{upquote.sty}{\usepackage{upquote}}{}
\IfFileExists{microtype.sty}{
  \usepackage[]{microtype}
  \UseMicrotypeSet[protrusion]{basicmath} 
}{}
\makeatletter
\@ifundefined{KOMAClassName}{
  \IfFileExists{parskip.sty}{%
    \usepackage{parskip}
  }{
    \setlength{\parindent}{0pt}
    \setlength{\parskip}{6pt plus 2pt minus 1pt}}
}{
  \KOMAoptions{parskip=half}}
\makeatother
\usepackage{xcolor}
\NewDocumentCommand\citeproctext{}{}
\NewDocumentCommand\citeproc{mm}{%
  \begingroup\def\citeproctext{#2}\cite{#1}\endgroup}
\makeatletter
 \let\@cite@ofmt\@firstofone
 \def\@biblabel#1{}
 \def\@cite#1#2{{#1\if@tempswa , #2\fi}}
\makeatother
\newlength{\cslhangindent}
\newlength{\csllabelwidth}
\newenvironment{CSLReferences}[2] 
 {\begin{list}{}{%
  \setlength{\itemindent}{0pt}
  \setlength{\leftmargin}{0pt}
  \setlength{\parsep}{0pt}
  \ifodd #1
   \setlength{\leftmargin}{\cslhangindent}
   \setlength{\itemindent}{-1\cslhangindent}
  \fi
  \setlength{\itemsep}{#2\baselineskip}}}
 {\end{list}}
\usepackage{calc}

\ifLuaTeX
\usepackage[bidi=basic]{babel}
\else
\usepackage[bidi=default]{babel}
\fi
\babelprovide[main,import]{american}

\def\languageshorthands#1{}
\ifLuaTeX
  \usepackage{selnolig}  
\fi
\IfFileExists{bookmark.sty}{\usepackage{bookmark}}{\usepackage{hyperref}}
\IfFileExists{xurl.sty}{\usepackage{xurl}}{} 
\hypersetup{
  pdftitle={STRAUSS: Sonification Tools \& Resources for Analysis Using
Sound Synthesis},
  pdfauthor={James W. Trayford, Samantha Youles, Chris Harrison, Rose
Shepherd, Nicolas Bonne},
  pdflang={en-US},
  colorlinks=true,
  linkcolor={Maroon},
  filecolor={Maroon},
  citecolor={Blue},
  urlcolor={Blue},
  pdfcreator={LaTeX via pandoc}}

\title{STRAUSS: Sonification Tools \& Resources for Analysis Using Sound
Synthesis}

\definecolor{c53baa1}{RGB}{83,186,161}
\definecolor{c202826}{RGB}{32,40,38}

\usepackage[affil-it]{authblk}
\usepackage{orcidlink}
\author[1%
  \ensuremath\mathparagraph]{James W. Trayford%
    \,\orcidlink{0000-0003-1530-1634}\,%
    }
\author[1%
  *%
  ]{Samantha Youles%
    }
\author[2%
  *%
  ]{Chris Harrison%
    }
\author[2%
  *%
  ]{Rose Shepherd%
    }
\author[1%
  *%
  ]{Nicolas Bonne%
    }

\affil[1]{Institute of Cosmology and Gravitation, University of
Portsmouth, Dennis Sciama Building, Burnaby Road, Portsmouth PO1 3FX,
UK%
  }
\affil[2]{School of Mathematics, Statistics and Physics, Newcastle
University, NE1 7RU, UK%
  }
\affil[$\mathparagraph$]{Corresponding author: %
}
\affil[*]{These authors contributed equally.}
\date{15 January 2024}

\begin{document}
\maketitle

\section{Summary}\label{summary}

\emph{Sonification}, or conveying data using non-verbal audio, is a
relatively niche but growing approach for presenting data across
multiple specialist domains including astronomy, climate science, and
beyond (\citeproc{ref-Lenzi21}{Lenzi et al., 2021};
\citeproc{ref-Lindborg23}{Lindborg et al., 2023};
\citeproc{ref-Zanella22}{Zanella et al., 2022}). The \texttt{strauss}
Python package aims to provide such a tool, which builds upon previous
approaches to provide a powerful means to explore different ways of
expressing data, with fine control over the output audio and its format.
\texttt{strauss} is a free, open source (FOSS) Python package, designed
to allow flexible and effective sonification to be straightforwardly
integrated into data workflows, in analogy to widely used visualisation
packages.

The remit of \texttt{strauss} is broad; it is intended to be able to
bridge between \emph{ad-hoc} solutions for sonifying very particular
datasets, and highly technical compositional and sound-design tools that
are not optimised for sonification, or may have a steep learning curve.
The code offers a range of approaches to sonification for a variety of
contexts (e.g.~science education, science communication, technical data
analysis, etc). To this end, \texttt{strauss} is packaged with a number
of examples of different sonification approaches, and preset
configurations to support a \emph{low-barrier, high-ceiling} approach.
\texttt{strauss} has been used to produce both educational resources
(\citeproc{ref-Harrison22}{Harrison et al., 2022}), and analysis tools
(\citeproc{ref-Trayford23b}{James W. Trayford et al., 2023}).

\section{Statement of need}\label{statement-of-need}

Sonification has great potential as a fundamental approach for
interfacing with data. This provides new perspectives on data that
complement visual approaches, as well as an accessible channel to data
for those who cannot access visual presentation, for example those who
are blind or visually impaired (BVI). As with the dominant approach of
data \emph{visualisation}, what can constitute sonification is very
broad, with different considerations for aspects such as the audience,
information content and aesthetics of the sonification. In order for
sonification to become more established and realise its potential as a
way of interfacing with data, accessible and flexible tools are needed
to make sonification intuitive and accessible to those who routinely
work with data.

Unlike data \emph{visualisation}, however, sonification of data is a far
less developed methodology, with a lack of widely adopted, cross-domain
tools and interfaces for those dealing with data to use. For example, in
the Python programming language, a number of packages exist for
visualising data, such as \texttt{matplotlib}
(\citeproc{ref-Hunter07}{Hunter, 2007}), \texttt{yt}
(\citeproc{ref-Turk11}{Turk et al., 2011}), \texttt{seaborn}
(\citeproc{ref-Waskom21}{Waskom, 2021}), or \texttt{plotly}
(\citeproc{ref-plotly}{Technologies Inc., 2015}). The lack of dedicated
and accessible tools for sonifying data is a barrier to exploring the
approach. Most solutions are piecemeal, using \emph{ad-hoc} tools to
parse data and map it to properties of sound as well as generating and
outputting sound in different formats. What's more, many of the tools
being repurposed to sonify data are proprietary and platform-dependent,
requiring paid-for licenses.

A number of effective python packages for data sonification have emerged
e.g.~\texttt{astronify} (\citeproc{ref-Brasseur23}{Brasseur et al.,
2023}) and \texttt{SonoUno} (\citeproc{ref-Casado24}{Casado et al.,
2024}). However these are typically feature-limited in how sonification
is produced, namely continuous pitch-mapped sonification.

Realising the potential of sonification may require a ``crowdsourced''
approach; via broad adoption of the technique and innovation in
sonification approaches driven by the scientific community. In this
context, we identify a need for Python package that:

\begin{itemize}
\item
  Provides a full pipeline from data to sonified audio that can be
  integrated into the workflows of scientists and data analysts
\item
  Is modular and enables complex, multi-variate sonification to be
  produced and fine tuned, while being simple enough to produce simple
  sonifications, for instance for novice users or in educational
  contexts
\item
  Is fully open-source and platform-independent, such that sonification
  is able to be integrated more broadly into scientific practice
\end{itemize}

\texttt{strauss} (\textbf{S}onification \textbf{T}ools and
\textbf{R}esources for \textbf{A}nalysis \textbf{U}sing \textbf{S}ound
\textbf{S}ynthesis) is a Python package for data sonification. The code
uses an object-oriented structure to provide a clear conceptual
framework for the different components of a sonification
(\citeproc{ref-Trayford23}{J. Trayford \& Harrison, 2023}).
\texttt{strauss} is designed to be used by analysts to fully sonify
their data in a way analogous to a plotting pipeline, to be analysed
independently or in conjunction with visuals.

By choosing the mapping between the variables being communicated and the
expressive properties of sound, \texttt{strauss} is designed to be a
\emph{low-barrier, high ceiling} tool; providing the means to make
diverse and highly customised sonifications with a high degree of
low-level control as the user becomes more experienced, while providing
a relatively simple path to produce sonifications quickly for a novice.
This is intended to provide the bridge between typical analysts who know
their data and the facets of it that they want to communicate, and the
sound experts or musicians that understand how to express data with
sound.

The \texttt{strauss} code minimises dependencies where possible,
implementing built-in signal generation and audio parsing and encoding
based on low-level python libraries like \texttt{numpy} and
\texttt{scipy}, as well as being tested on multiple platforms
(\emph{MacOS, Linux, Windows}). This also provides means to interface
with commonly used audio formats, such as \emph{``Soundfont''} files
(\texttt{.sf2}) to open a range of possibilities for representing data
using different instruments. This helps to ensure that the free and open
source status of \texttt{strauss} is maintained and does not require
difficult to install or proprietary software, that may themselves have
steep learning curves.

Towards our \emph{low barrier} aspirations for \texttt{strauss}, a
Tutorial-driven development (TDD) approach is used where each major
feature should have an associated tutorial example, which are packaged
with the code, in both Python Notebook (\texttt{examples/.ipynb}) and
Python script (\texttt{examples/*.py}) formats. \texttt{strauss}
provides tools for inline playback of sound in both formats. In
addition, a further collection of modified examples is provided
specifically for the \emph{Google Colab} platform
(\texttt{examples/colab/*.ipynb}), allowing examples to be run fully
in-browser without requiring a local install of the code.

Offering both formats for examples in \texttt{strauss} allows us to
exploit the interactivity and low technical threshold needed to use
notebooks, while also having the BVI accessibility of the raw-text
scripts, given the difficulties of using notebooks with screen readers
(\citeproc{ref-Trayford24}{J. W. Trayford et al., 2024}).
\texttt{strauss} is provided with extensive documentation, maintained
and hosted via
\href{https://strauss.readthedocs.io/en/latest/}{\emph{Read The Docs}}.

\section{Acknowledgements}\label{acknowledgements}

JWT acknowledges support via the STFC Early Stage Research \&
Development Grant, reference ST/X004651/1, CMH acknowledge funding from
an United Kingdom Research and Innovation grant (code: MR/V022830/1). RS
is supported by a studentship from an STFC Centre of Doctoral Training
in Data Intensive Science (code: ST/W006790/1).

\section*{References}\label{references}
\addcontentsline{toc}{section}{References}

\phantomsection\label{refs}

\end{document}